# Modèles biomécaniques pour simuler les conséquences des chirurgies maxillo-faciales

# Biomechanical models to simulate consequences of maxillofacial surgery

## Modèles biomécaniques et chirurgie maxillo-faciale

*Yohan Payan[1], Matthieu Chabanas[1], Xavier Pelorson[2], Coriandre Vilain[2], Patrick Levy[1,3], Vincent Luboz[1] & Pascal Perrier[2]*

[1]Laboratoire TIMC - Faculté de Médecine - Domaine de la Merci - 38706 La Tronche cedex - France
[2]Institut de la Communication Parlée – 46 Av. Felix Viallet - 38031 Grenoble cedex - France
[3]Laboratoire du Sommeil - CHU de Grenoble - BP 217 - 38043 Grenoble cedex 9 - France

**Auteur correspondant**: Yohan Payan, Laboratoire TIMC - Faculté de Médecine – Institut Albert Bonniot - 38706 La Tronche cedex - France, tel: +33 (0)476 54 95 22, fax: +33 (0)476 54 95 55, Yohan.Payan@imag.fr

**Résumé:**

Ce papier introduit deux modèles biomécaniques développés dans un cadre de chirurgie maxillo-faciale assistée par ordinateur. Après un rapide résumé des outils de modélisation utilisés, ces modèles biomécaniques sont présentés puis leurs résultats discutés.

Le premier modèle touche la chirurgie othognathique et vise à prédire les conséquences des ostéotomies maxillaires et mandibulaires. Pour cela, un modèle tridimensionnel générique de la face est automatiquement adapté (par recalage élastique) à la morphologie du patient. Ce modèle est alors utilisé pour simuler de manière qualitative les conséquences d'ostéotomies mandibulaires.

Le second modèle vise l'étude du Syndrome d'Apnées du Sommeil. Il s'agit de développer un modèle réaliste de l'interaction entre flux d'air et parois déformables des Voies Aériennes Supérieures. Des simulations numériques sont présentées et qualitativement comparées à des mesures faites sur patient, au cours de cycles respiratoires.

**Abstract:**

This paper presents the biomechanical Finite Element models that have been developed in the framework of the computer-assisted maxillofacial surgery. After a brief overview of the continuous elastic modeling method, two models are introduced and their use for computer-assisted applications discussed.

The first model deals with orthognathic surgery and aims at predicting the facial consequences of maxillary and mandibular osteotomies. For this, a generic three-dimensional model of the face is automatically adapted to the morphology of the patient by the mean of elastic registration. Qualitative simulations of the consequences of an osteotomy of the mandible can thus be provided.

The second model addresses the Sleep Apnea Syndrome. Its aim is to develop a complete modeling of the interaction between airflow and upper airways walls during respiration. Dynamical simulations of the interaction during a respiratory cycle are computed and compared with observed phenomena.

**Version abrégée**

# I. INTRODUCTION

Le programme GMCAO (Gestes Médico-Chirurgicaux Assistés par Ordinateur) du laboratoire TIMC de l'IMAG vise à assister les chirurgiens dans la réalisation de gestes diagnostiques ou thérapeutiques les plus précis et les moins invasifs possibles. Dans ce cadre, les chercheurs du laboratoire sont amenés à développer des modèles d'organes humains (on parle de modèles « biomécaniques » ou de modèles « physiques »), et à utiliser ces modèles pour simuler numériquement les conséquences de certains actes chirurgicaux.

Les structures osseuses ont été les premières à être ainsi modélisées, essentiellement du fait de leur rigidité et donc de la relative facilité à modéliser numériquement de telles structures. C'est plus récemment que les chercheurs ont abordé la modélisation des structures anatomiques molles dans des cadres de chirurgie assistée par ordinateur. Ce papier focalise sur l'intérêt de la mise place de tels outils de modélisation, en prenant pour illustration le domaine de la chirurgie maxillo-faciale assistée par ordinateur. Après une brève introduction sur les outils numériques de modélisation utilisés, deux applications seront plus particulièrement abordées : la chirurgie orthognathique (avec le développement d'un modèle biomécanique de la face) et le Syndrome d'Apnées du Sommeil (avec la mise au point de modèles physiques d'interaction flux d'air / parois déformables).

# II. METHODE DE MODELISATION DES TISSUS MOUS

L'outil mathématique qui semble le plus propice à une modélisation des propriétés d'élasticité continue des structures anatomiques molles est sans doute la *Méthode des Éléments Finis* (*MEF*). Cette formulation a en effet le double avantage de rester fidèle aux propriétés élastiques et continues (puisqu'elle consiste, en fait, en une discrétisation des équations d'élasticité), tout en permettant, grâce à la notion d'*élément*, de définir différentes zones, avec des propriétés mécaniques distinctes. Les points clefs de cette modélisation par éléments finis peuvent se résumer par :
- un partage du domaine en *éléments finis* standards délimités par un certain nombre de *noeuds*.
- les équations d'élasticité continue sont discrétisées en chacun des noeuds du domaine : les déplacements, vitesses et accélérations de ces noeuds, sont alors les inconnues du système, tandis que les variables cinématiques de chacun des points d'un élément sont obtenues après interpolation entre les noeuds.

- les propriétés physiques du matériau à modéliser, sont prises en compte par l'intermédiaire de deux *constantes d'élasticité* : le module d'Young $E$ (caractérisant la raideur du matériau) et le taux de Poisson $\nu$ (caractérisant la plus ou moins grande compressibilité du matériau).
- la méthode offre alors la possibilité d'associer à chaque élément du domaine des propriétés physiques distinctes (avec, par exemple, des éléments ayant le comportement des tissus passifs, tandis que d'autres peuvent simuler les propriétés de fibres musculaires).

## III. MODELISATION BIOMECANIQUE DE LA FACE POUR LA CHIRURGIE ORTHOGNATHIQUE

### III.1 Contexte

La chirurgie orthognathique vise à corriger les dysmorphoses maxillo-mandibulaires, par découpe des branches mandibulaires et/ou de la partie supérieure du maxillaire. Les implications de cette chirurgie sont à la fois fonctionnelles, pour la normalisation de l'occlusion dentaire, et esthétiques, pour le rééquilibrage de la face. Lors des deux projets européens IGOS I et II (*Image Guided Orthopaedic Surgery*), le laboratoire TIMC a développé un simulateur pour l'aide au repositionnement des structures osseuses mandibulaires et maxillaires. L'idée sous-jacente consistait à fournir au chirurgien une reconstruction 3D de la morphologie du crâne du patient (à partir d'un examen tomo-densitométrique) et à y ajouter des outils de planification des découpes osseuses qui suivent des protocoles céphalométriques existant. De plus, les contraintes en terme d'occlusion dentaire, telles qu'elles ont été définies par l'orthodontiste, ont été intégrées au planning final. La suite logique de ce travail de planification concerne la simulation des conséquences de ces découpes osseuses sur l'esthétique du visage, ainsi que sur les fonctionnalités de la partie basse du visage (mimiques faciales, mastication, production de parole). C'est ainsi qu'un modèle biomécanique des tissus mous peauciers a dû être développé et utilisé en simulation.

### III.2. Modélisation biomécanique

Élaborer un modèle biomécanique donné, selon une morphologie particulière à un patient, est une tâche longue et fastidieuse que nous ne souhaitons pas renouveler pour chaque patient. C'est la raison pour laquelle notre méthodologie de modélisation a été la suivante :
1. Un modèle "standard" de la face (on peut parler alors d'« atlas ») a été élaboré manuellement, avec une spécificité dans la définition géométrique du maillage 3D.
2. A partir de données d'imagerie scanner du patient, une méthode de recalage élastique local (l'algorithme "Mesh-Matching") déforme le modèle "standard" de la face vers la géométrie du patient, afin de générer automatiquement un nouveau modèle biomécanique adapté à la morphologie du visage du patient.

Le modèle "standard" de la face est basé sur un maillage volumique multicouches représentant les différentes épaisseurs du visage : épiderme, derme et hypoderme. Les équations d'élasticité linéaire sont discrétisées suivant le formalisme de la méthode des Eléments Finis. Les propriétés mécaniques (module d'Young, taux de Poisson) sont fixées en cohérence avec les études mécaniques réalisées sur tissu peaucier humain. La peau et les tissus graisseux sont considérés comme isotropes et incompressibles. Des structures indépendantes, représentant les principaux muscles faciaux, sont ensuite insérées dans le maillage 3D, avec des propriétés mécaniques différentes : orthotropie suivant les directions principales des fibres musculaires, élasticité dépendant de l'activité du muscle.

Une fois déformé pour approcher la morphologie de la face du patient, mesurée sur un examen scanner, ce modèle biomécanique peut alors être utilisé en simulation, afin de qualitativement observer les conséquences esthétiques et fonctionnelles des repositionnements des structures osseuses.

## IV. MODELISATION BIOMECANIQUE DES VOIES AERIENNES SUPERIEURES POUR LE SYNDROME D'APNEES DU SOMMEIL

### IV.1 Contexte

Les Voies Aériennes Supérieures (VAS) sont délimitées d'une part par des structures mécaniquement rigides (telles que le palais dur ou les dents), et d'autre part par des tissus mous (langue, pharynx, vélum, cordes vocales), déformables sous l'action des forces musculaires ou de forces externes telles que celle qui sont générés par l'écoulement d'air. Dans certaines conditions nocturnes, les phénomènes d'interaction fluide-parois génèrent une auto oscillation des VAS (comme c'est le cas pour le ronflement ) mais peuvent aussi conduire jusqu'à une occlusion partielle (« hypopnée ») ou totale (« apnée ») des VAS. La survenue de plus de 5 apnées ou 10 apnées + hypopnées par heure de sommeil et l'existence de symptômes nocturnes (le ronflement essentiellement) et diurnes (la somnolence diurne excessive, surtout) définissent le Syndrome d'Apnées du Sommeil (SAS). Cette maladie touche 4% des hommes et 2% des femmes entre 30 et 60 ans, et ses conséquences peuvent être parfois extrêmement gênantes, du fait du sommeil perturbé et de la fatigue diurne associée. La chirurgie est une des réponses cliniques apportées à cette maladie, avec le plus souvent une ostéotomie mandibulaire visant à élargir le calibre pharyngé par une avancée de l'articulateur lingual. Si les résultats de cette chirurgie sont le plus souvent satisfaisant, il reste extrêmement difficile de prédire a priori la probabilité de réussite de ce geste chirurgical, et ceci essentiellement du fait de l'extrême complexité des phénomènes physiques mis en jeu dans le SAS. Notre travail vise donc à tenter de mieux comprendre ces phénomènes physiques, en recourant notamment aux outils de modélisation numérique du couplage entre le flux d'air et les parois déformables des VAS.

**IV.2 Modélisation physique du SAS**

D'un point de vue physique, le phénomène d'apnée du sommeil peut être analysé comme le résultat d'une interaction spectaculaire entre l'écoulement d'air issu des poumons et les parois du conduit vocal. Il est donc évident que pour pouvoir comprendre et modéliser ce phénomène il est nécessaire de disposer d'une part d'une description réaliste de l'écoulement mais aussi, d'autre part, d'un modèle destiné à reproduire les caractéristiques biomécaniques du pharynx et de la langue.

Du point de vue aérodynamique, la complexité du problème s'explique en grande partie parce que les phénomènes rencontrés dépassent largement le cadre des hypothèses usuelles. En particulier, lors de l'apparition d'un collapsus dans le conduit vocal, les écoulements sont essentiellement dominés par les effets visqueux et non stationnaires qui sont décrits de manière trop simplifiée par les modèles physiques classiquement présentés dans la littérature.

Du point de vue biomécanique, la prise en compte réaliste de cette interaction nécessite une description suffisamment fine de l'élasticité des tissus des parois des VAS, dans l'objectif de prédire correctement l'amplitude et la nature des déformations ainsi occasionnées.

Ces questions ont déjà fait l'objet de travaux à TIMC et au Laboratoire du Sommeil du CHU de Grenoble dans le cadre de l'étude des pathologies des voies aériennes supérieures, et à l'Institut de la Communication Parlée (ICP) de Grenoble en ce qui concerne la modélisation des processus de production de la parole. A l'ICP, les travaux ont permis de proposer une description plus réaliste des phénomènes aérodynamiques les plus importants tels que le décollement tourbillonnaire de l'écoulement. À TIMC et à l'ICP un modèle biomécanique de la langue a été développé qui prend en compte les principaux muscles agissant sur la forme et la position de la langue dans le plan medio-sagittal de la tête, ainsi qu'un modèle simplifié des structures vélaires. Au Laboratoire du Sommeil, enfin, de nombreuses données physiologiques ainsi que des mesures aérodynamiques ont été acquises sur des patients atteints du syndrome d'apnée du sommeil

Les premières simulations de couplage flux / parois montrent que les phénomènes classiques de réduction de débit observés lors d'hypopnées peuvent être qualitativement obtenus par simulation numérique. D'autre part, l'influence de la rigidité des tissus sur la probabilité d'apparition du SAS (cf. aussi les techniques d'uvulo-plastie) est également retrouvée de manière qualitative par simulation.

## IV. CONCLUSION

Même si leurs résultats doivent toujours être observés avec précaution, les modèles physiques (biomécanique des tissus mous, aérodynamique de l'écoulement) peuvent apporter, de notre point de vue, des compléments intéressants à la planification des actes chirurgicaux. Nous avons tenté dans ce papier d'en présenter les premières lignes et d'ébaucher des protocoles de chirurgie assistée par ordinateur dans les domaines du Syndrome d'Apnées du Sommeil et de la chirurgie orthognathique.

# BIOMECHANICAL MODELS TO SIMULATE CONSEQUENCES OF MAXILLOFACIAL SURGERY

## I. INTRODUCTION

The Computer-Assisted Medical Imaging (CAMI) group of TIMC laboratory aims at assisting surgeons for the realization of diagnostic and therapeutic gestures that have to be safe and precise. In this framework, numerical models of human anatomical structures have been developed (the terms "physical" or "biomechanical" models are often used) to help surgeons planning their surgical acts. First, models are used to replicate and to understand the behavior of the human mechanical structures. Then, once the models are validated, they can be used to numerically simulate the consequences of the surgery on these structures. This last step is of course the main objective for a CAMI modeling application, but it should not occult the huge amount of work required to build up and to validate a model, as well as the fundamental role of a model in the understanding of physical phenomena. The predictions made from a model are not a perfect representation of the real impact of the surgery, since a biomechanical model is always an approximation of the complex physical entities that define any anatomical structures. One must be aware of these limits inherent to any modeling approach, in analyzing the results. Knowing exactly, which ones they are, and to which extent they can distort the reality, allows us to properly interpret the simulations.

The first modeled anatomical structures were bones, mainly because they can be considered as rigid, which allows to use quite simple modeling framework. More recently, biomechanical models were developed to replicate and to understand the behavior of soft tissues and organs such as the heart, the liver, the prostate or the kidney. The modeling of this kind of tissues requires a much stronger mathematical basis, in order to discretize the partial differential equations that govern the continuum mechanics. This paper focuses on two models built up for computer-aided plastic and maxillofacial surgery. The first one concerns the biomechanical modeling of the human face. This model is the much "advanced" one, as it will soon lead, from our point of view, to a clinical routine use. The second example concerns the sleep apnea syndrome. At the present stage of the work, this model is mainly used to understand the complex physical phenomena that lead to the collapsus of the upper airways.

The modeling framework used to describe elastic deformations of soft tissues is introduced in the first part of this paper. Then, a biomechanical model of the face for computer assisted orthognathic surgery is presented. Finally, the sleep apnea syndrome is introduced and the first models developed to understand this phenomenon are described.

## II. SOFT TISSUES MODELING METHOD

The mathematical tool chosen to describe soft tissues deformation is the *Finite Element Method* (FEM). This method allows a precise description of the continuous and elastic properties of a body. Moreover, it is possible, via the notion of *element*, to attribute specific biomechanical properties to individual regions of the structure. The mathematical formulation of the FEM is detailed in [1] and [2]. General principles of this method can be summarized as follows:

- the body is divided into standard *finite elements*, delimited by *nodes* (the set of elements is called *mesh)*;
- the elasticity equations are computed at each node: the nodes displacements, velocities and accelerations, are then the unknown variables of the system, while kinematics parameters inside each element are computed by interpolating between the nodes ;
- the physical properties of the body are taken into account through two elastic constants: the *Young modulus E* and the *Poisson ratio ν*.

Those two constants have a crucial influence in the behavior of the Finite Element structure; their respective values have therefore to be chosen carefully.

### *The Young modulus E*

The Young modulus $E$ measures the way the body resists to deformations induced by external forces. It has the dimension of a pressure force: a quite-rigid body has a high modulus, while a malleable one gets a lower one. Figure 1 shows, for a simple square FE structure laid on a horizontal plane (i.e. with constraints of no-displacement applied to the lower nodes of the square mesh), the deformations induced by vertical forces for two different values of the Young modulus. The Young modulus of the left structure is ten times larger than for the right one. Note here that the Finite Element codes that were developed for this example assume no displacement in the transverse direction (*the plane strain hypothesis*) as well as a small deformation framework (with relative deformations of the structure that do not exceed 10%). For those reasons, the results presented above should be considered as very qualitative ones, which give an idea of the main trends of the soft body deformation, and not as true representation of the reality.

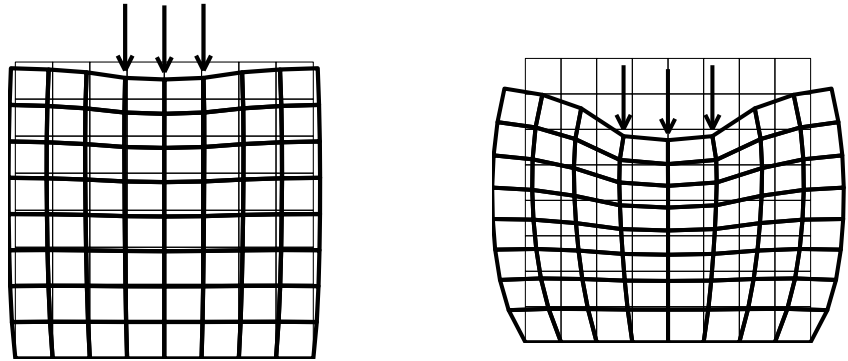

Figure 1: Young modulus influence: E = 60 kPa (left) and E = 6 kPa (right)
Poisson's ratio $\nu$ is fixed to 0.49

*The Poisson ratio*

The Poisson ratio $\nu$ describes the way a pressure force, applied in a specific direction, induces deformations in other directions. This parameter is sometimes associated with the way the structure accepts variations of its own volume. For example, a ratio close to zero means that the structure deforms along a dimension, without any change along the other dimensions of the system. On the contrary, in the case of a three-dimensional structure, a Poisson ratio close to the value of 0.5 represents a quasi-uncompressible behavior (at the second order level), i.e. with volume conservation (see [1] for details). Figure 2 plots the deformations induced by the pressure forces for two different Poisson ratios.

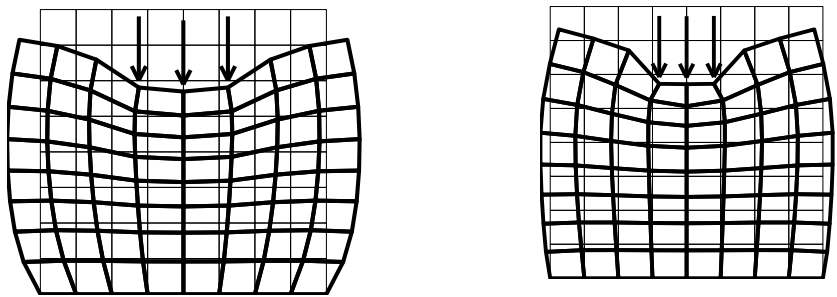

Figure 2: Poisson ratio influence: $\nu$ = 0.49 (left) and $\nu$ = 0.2 (right)
Young modulus E is fixed to 6 kPa

## III. BIOMECHANICAL MODEL OF THE FACE FOR ORTHOGNATHIC SURGERY

### III.1. Context

Orthognathic surgery deals with skeletal deformities of the lower part of the skull. The correction of these deformities requires the cutting and repositioning of bones by osteotomies of the mandible and/or the maxilla. In the course of the European IGOS project (*Image Guided Orthopaedic Surgery*), the TIMC/CAMI group developed a 3D simulator to plan bone-repositioning in maxillofacial surgery [3]. This simulator integrates 3D cephalometric constraints for bony structures repositioning and orthodontic constraints for teeth occlusion. Following the diagnosis for repositioning of the bone structures provided by this simulator, biomechanics of the patient facial soft tissues has to be taken into account, in order to predict the « external aspect » of the patient face after bones displacements. This complex prediction requires the elaboration of a complete 3D biomechanical model of the patient face, integrating the passive skin tissues and the active muscular structures that deform the face.

### III.2. Methodology

Building up a 3D biomechanical model of the patient face is a complex task, highly demanding from a modeling point of view. Therefore, to avoid the prohibitive amount of manual work required to generate a model for each patient, the following strategy was adopted:

3. A generic face model, or "atlas", was manually elaborated including a specific geometrical design of the mesh.
4. Starting from patient data (Computer Tomography data in this case), a non-rigid registration algorithm, the "Mesh-Matching" [4], was used to adapt the generic model to the patient morphology, thus generating a Finite Element model of the patient face.

### III.3. A generic 3D biomechanical model of the face

Facial skin has a layered structure composed of the epidermis (0.1 mm thick), dermis (0.5 - 3.5 mm thick) and hypodermis (fatty tissues connected to the skull). Facial muscles are inserted between those skin layers and the skull. As many orthognathic interventions concern the mandible and the maxilla, the modeling was mainly dedicated to the lower part of the face, with a special emphasis on soft tissues surrounding human lips.

In that perspective, a 3D generic mesh of the human face was manually designed [5]. It is made of two layers of hexahedral and wedge elements (figure 3).

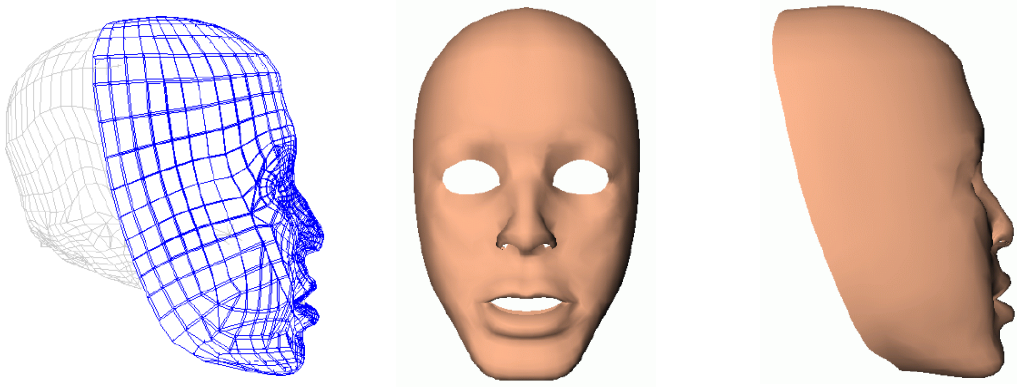

Figure 3: Finite Element mesh (left) and rendered views (middle and right)

Finite Element Method is then used to discretize the linear elasticity equations at the nodes of this 3D mesh. For this, the commercial ANSYS™ software was used, mainly because of the possibilities this Finite Element package offers in terms of large deformations and hyper-elasticity modeling framework. Biomechanical parameters (Young modulus and Poisson ratio) are chosen to replicate measurements made on human skin tissues. A value close to 0.5 is therefore chosen for the Poisson ratio, in order to model the quasi-incompressibility of tissues (which are mainly composed of water). Concerning the Young modulus values, distinctions were made between elements that were associated with passive skin tissues, and elements labeled as active "muscular" elements. Based on Fung's measurements [6], a value of 15 kPa is used for passive skin tissues. Measurements reported by Duck [7] were used for the mechanical characteristics of "active" muscular elements, with a transverse isotropic stress/strain relationship. This requires two parameters for stiffness: a Young modulus $E_{fibers}$ in the main direction of the fibers, and another modulus $E_{ortho}$ in the orthogonal directions. Following Duck's measurements, a stiffness of 6.2 kPa is used for both Young moduli at rest, while a 110 kPa value is given to $E_{fibers}$ when muscle activation is maximum. $E_{fibers}$ values therefore linearly increase with muscular activation, while $E_{ortho}$ remains constant.

The main facial muscles surrounding the lips were taken into account during the manual design of the 3D mesh. A set of elements inside this mesh were therefore labeled as muscular elements and associated with the following muscles: Buccinator, Risorius, Zygomaticus major and minor, Depressor anguli oris and Orbicularis oris (see [5] for details). Figure 4 plots the face deformations induced by the activation of the Zygomaticus major muscle (with a 1 Newton force value) in the model.

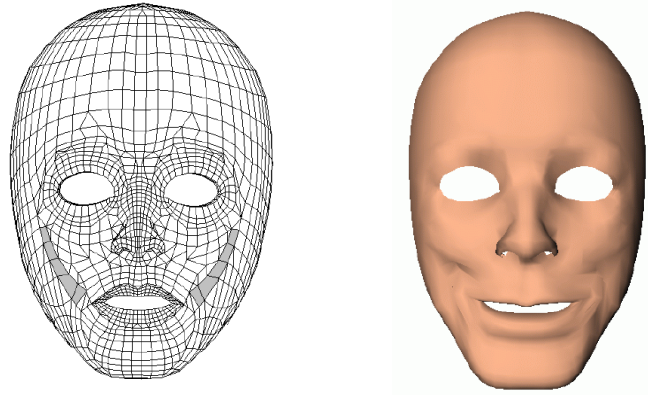

Figure 4 : Zygomaticus muscle elements (left) and their induced facial deformations (right)

### III.4. Adaptation of the generic model to the patient morphology

Data were extracted from a complete CT exam of the patient head. Two datasets of 3D points were automatically extracted: points located on the external surface of the skull (marching cubes algorithm, [8]) and points located on the patient's skin surface (automatic segmentation procedure). The Mesh-Matching algorithm [4] was then used to match the nodes of the generic mesh to these two datasets. The external nodes of the generic mesh were transformed to match the points segmented from the patient's skin surface. In the same way, a non-rigid transformation was calculated to project the internal nodes of the generic mesh onto the points located on the external surface of the patient's skull. By connecting together the nodes transformed by the Mesh-Matching algorithm, a new 3D mesh was provided for Finite Element Analysis adapted to the patient morphology. Figure 5 shows the automatically generated model of a first patient.

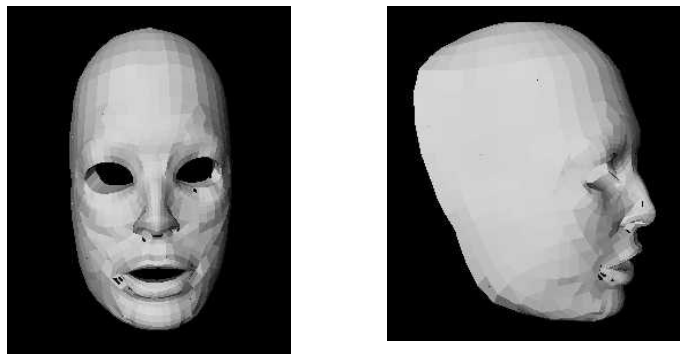

Figure 5: Front (left) and profile (right) views of the FE model of the patient

Figure 6 shows qualitative results for the simulation of an osteotomy of the patient mandible. A forward 5 mm displacement of the mandible was simulated. Note that relative deformations of the face, in the chin and lips regions can be observed.

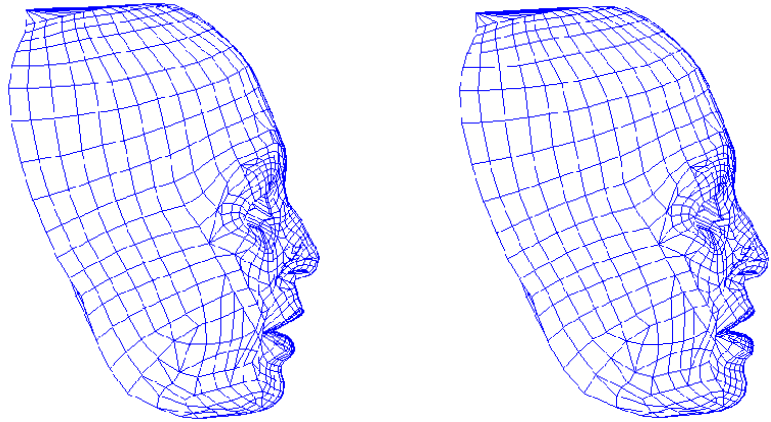

Figure 6: Simulation of a forward displacement of the patient mandible

## IV. BIOMECHANICAL MODELS FOR THE SLEEP APNEA SYNDROME

Sleep Apnea Syndrome (SAS) is defined as a partial or total closure of the patient upper airways during sleep. The term "collapsus" (or collapse) is used to describe this closure. From a fluid mechanical point of view, this collapse can be understood as a spectacular example of fluid-walls interaction. Indeed, the upper airways are delimited in their largest part by soft tissues having different geometrical and mechanical properties: velum, tongue and pharyngeal walls. Airway closure during SAS comes from the interaction between these soft tissues and the inspiratory flow.

The aim of this work is to understand the physical phenomena at the origin of the collapsus and the metamorphosis in inspiratory flow pattern that has been reported during SAS. Indeed, a full comprehension of the physical conditions allowing this phenomenon is a prerequisite to be able to help in the planning of the surgical acts that can be prescribed for the patients. Surgical techniques used for the treatment of the SAS can either reduce the volume of the tongue or stiff the velum, or try to have a more global and progressive action on the entire upper airways.

The work presented here focuses on a simple but coherent model of fluid-walls interactions. The equations governing the airflow inside a constriction are first presented. Then, a biomechanical model of the velum (or soft palate) will be described and its deformations and interactions with the airflow detailed. Finally, simulations provided by the coupled models will be discussed and compared to measurements.

### IV.1. Fluid mechanical aspects

To illustrate our approach, figure 7 presents in a simple way a constriction inside the upper airways.

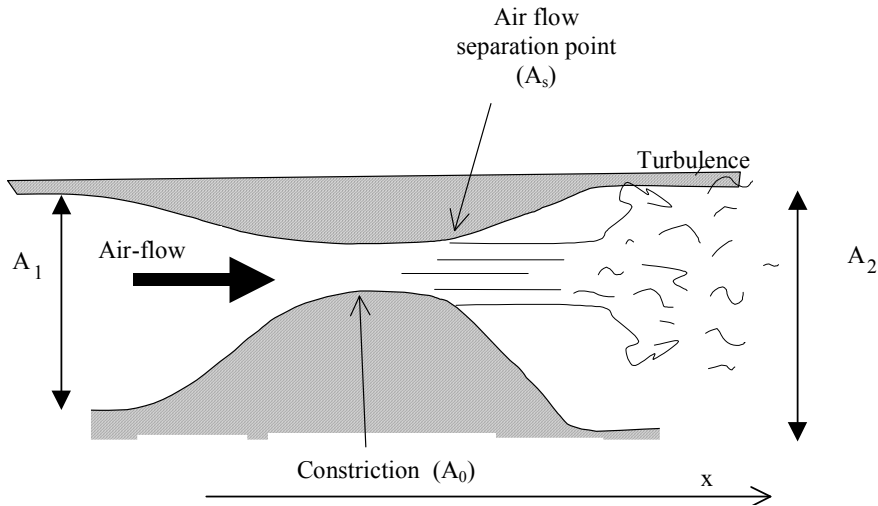

**Figure 7: Schematic Illustration for airflow inside the constriction.**

An exact analytical solution for the flow through such a constriction is not available. Further, full numerical simulations of the unsteady three dimensional flow through a deformable structure are still, at present time, impossible even using the recent numerical codes and using powerful computers [9]. For these reasons, and also because the aim of this paper is to provide a qualitative description of a sleep apnea, we will use in the following a very simplified flow theory based on the following assumptions:

- As the airflow velocity in the upper airways is, in general, much smaller than the speed of sound (low Mach number flow), it can be assumed that the flow is locally incompressible.
- It can be reasonably assumed that the time needed for the constriction to collapse (of order of a second) is large compared with typical flow convection times (the time needed for the flow to pass the constriction is of order of a few milliseconds). Therefore, it will be assumed that the flow is quasi-steady (low Strouhal number flow).

The principle of mass-conservation thus yields the following relationship:

$$\Phi = \text{constant} \qquad (1)$$

where $\Phi = v.A$ is the volume flow velocity, $v$ and $A$ are respectively the (local) flow velocity and vocal tract area. As a third and last assumption, we now consider that all viscous effects can be neglected. This assumption can be rationalized partially by considering that typical Reynolds numbers involved are of order of 1000 and thus that viscous forces are negligible compared with convective ones. This leads to the well-known one-dimensional equation (Bernoulli law):

$$p + \frac{1}{2}\rho v^2 = \text{constant} \qquad (2)$$

where $p$ is the local pressure and $\rho$ the (constant) air density

Equations (1) and (2) must be corrected in order to take into account a spectacular viscous effect: flow separation. Indeed, it is expected that the strongest pressure losses are due to the phenomenon of flow separation at the outlet of the constriction. This phenomenon is due to the presence of a strong adverse pressure gradient that causes the flow to decelerate so rapidly that it separates from the walls to form a free jet. Very strong pressure losses, due to the appearance of turbulence downstream of the constriction, are associated with flow separation. As a matter of fact, the pressure recovery past the flow separation point is so small that it can in general be neglected. In the following, we consider that the flow separates from the walls of the constriction at the point where the area reaches 1.2 times the minimum area $A0$ (see figure 7). This approximated value was empirically proposed and constitutes an acceptable approximation of the phenomena [10].

To summarize, for a given pressure drop ($p1-p2$), and for a given geometry of the constriction, the volume flow velocity $\Phi$ is:

$$\phi = A_s \sqrt{\frac{2(p_1 - p_2)}{\rho}} = 1.2 A_0 \sqrt{\frac{2(p_1 - p_2)}{\rho}} \quad (3)$$

and the pressure distribution $p(x)$ within the constriction is predicted by:

$$p(x) = p_1 + \frac{1}{2}\rho\phi^2 \left( \frac{1}{A_1^2} - \frac{1}{A(x)^2} \right) \quad (4)$$

where $A(x)$ is the transversal area at the $x$ abscissa (figure 7).

Thus, the force exerted by the airflow onto the walls of the constriction can be computed by integrating the pressure along the $x$ axis up to the flow separation point. This force induces a deformation of the upper airways soft tissues, thus modifying the airways geometry, and therefore changing the pressure distribution along the airways. Next part describes the model that was developed to simulate this phenomenon.

### IV.2. A biomechanical model of the velum

The soft tissues that constitute the tongue, the soft palate (including the velum) and the pharyngeal walls are partly responsible for the SAS as their deformations can even lead to a total closure of the upper airways. In a first step, we only focused on the velo-pharyngeal region, at the intersection between the nasal and the oral cavity. In this perspective, a continuous model of the velum was elaborated and defined through the Finite Element Method. The codes that were developed for this model assume no displacement in the transverse direction (*the plane strain hypothesis*) as well as a small deformation framework Again, for quasi-incompressibility reasons, a value close to 0.5 was chosen for the Poisson ratio. A 10 kPa value was taken for the Young modulus, which seems coherent with values reported for tongue [11] and vocal folds [12]. The geometry of the model was extracted from a single midsagittal radiography of a patient (figure 8).

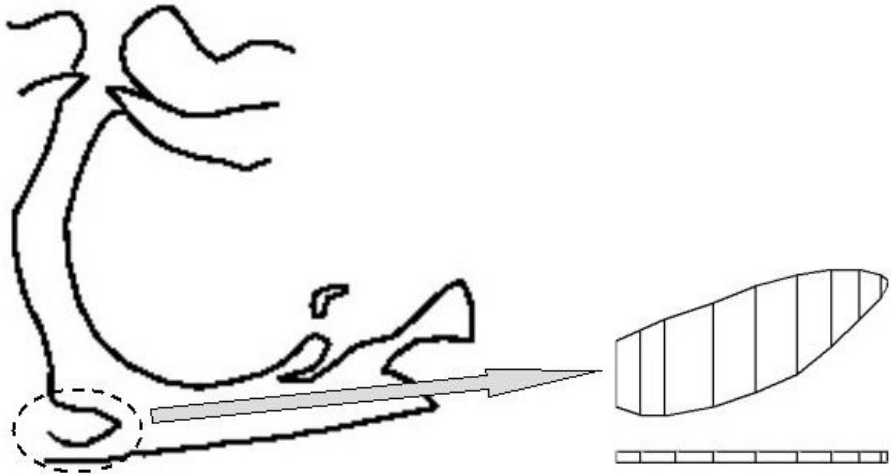

Figure 8 : Midsagittal view of the upper airways (left) and FE model of the velum (right).

The upper part of the model represents the velar tissues. Geometrical values were coherent with values reported in the literature [13] [14]: thickness of the velum varies from one extremity to the other (with a mean of 5-mm) while the total length is of order of 30-mm.

For the current simulations, only the velum deformations were taken into account, as the pharyngeal walls (the lower part of the model on figure 8) were supposed to be rigid during FE computations. The two points located onto the left part of the velum were also considered as fixed in order to model the velar attachment to the hard palate. Last, simulations are limited to the midsagittal plane, but for the computation of pressure forces, a 30-mm value was taken for the velar thickness in the frontal plane.

### IV.3. Coupling the airflow pressure forces and the velar model deformations

An iterative process governed the coupling between the airflow pressure forces computation and the deformations of the FE model of the velum. An adaptive Runge Kutta algorithm was used to solve the dynamical equations that govern the deformations of the model and its coupling with the airflow.

### IV.4. Simulations of airflow limitation

In order to simulate a respiratory cycle, the pressure difference ($p1$-$p2$) was approximated by a sinusoid: $p_1 - p_2 = p_{max} * \sin(4 * \pi * t)$. Figure 9 shows this pressure command for half a period and with an 800 Pa maximal value.

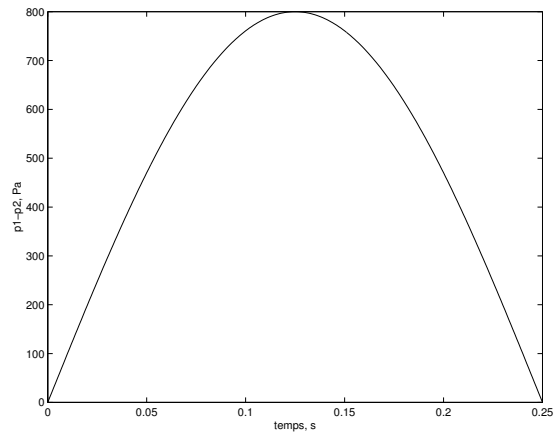

**Figure 9: Command of the model: maximal pressure difference $p_{max}$ = 800 Pa.**

For computation-time reasons, the time scale was divided by a factor of ten. Figure 10 shows results of the simulations for the command depicted in figure 9. A clear reduction of the constriction can be observed, thus simulating a hypo-apnea [15].

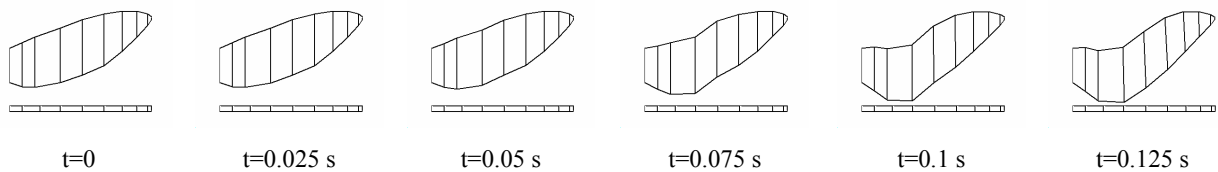

| t=0 | t=0.025 s | t=0.05 s | t=0.075 s | t=0.1 s | t=0.125 s |

**Figure 10: Deformations of the model coupled with airflow:
from initial (left) to final (right) positions.**

These results are consistent with the volume flow velocity limitation observed on patient suffering from SAS. Figure 11 plots the simulated volume flow velocity for three values of $p_{max}$. A flow limitation is clearly identifiable at $p_{max}$ = 800 Pa but tends to disappear when $p_{max}$ decreases. This result is qualitatively consistent with the use for the nasal Continuous Positive Airway Pressure [16], which in case of flow-limitation syndrome is reducing the pressure drop.

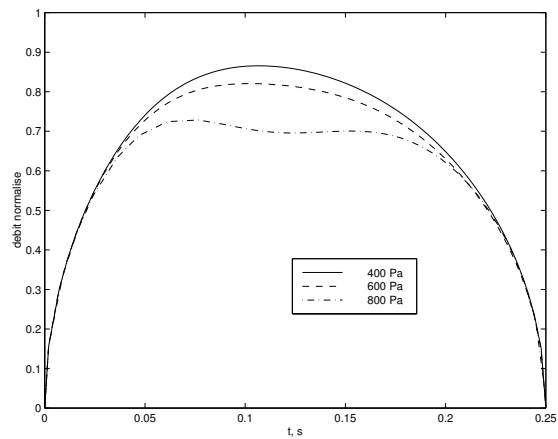

**Figure 11: Influence of the pressure drop onto the volume flow velocity.**

### IV.5. Influence of the velar tissues stiffness

Figure 12 presents the simulated volume flow velocity for three different values of the Young modulus, that is to say for three different stiffness of the velum. As can be clearly noticed here, an increase of the stiffness value of the velum tends to reduce the hypo-apnea phenomena. This result is consistent with some surgery techniques that try to modify mechanical properties of the velum by burning tissues, thus increasing their stiffness.

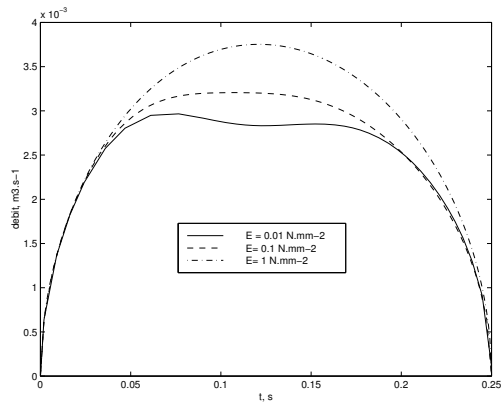

**Figure 12: Influence of the velum stiffness onto the volume flow velocity.**

## V. CONCLUSION

This paper presented two biomechanical models developed in the framework of plastic and maxillofacial surgery. Both models are based on the Finite Element Method to solve the partial differential equations that govern the continuous elasticity laws. The biomechanical FE model of the human face aims at predicting the facial aesthetic and functional consequences of orthognathic surgery. A first clue was given in our ability to build a model adapted to the geometry of a patient and to carry out simulations with this model. In an even more qualitative framework, first simulations were provided for the fluid-wall interactions during respiration. The first results seem to qualitatively validate the behavior of the model as observed phenomena were numerically simulated. Contrary to the model of the face, this airflow/velum model needs some improvements before it can be used to predict consequences of surgery. Further works will include the modeling of the other structures of the upper airways (tongue and pharyngeal walls) [17], a refinement of the flow model, with, in particular, an evaluation of the adequacy of the quasi-steady inviscid assumptions during the initiation of a collapse or when permanence of turbulence within the constriction is expected.

ACKNOWLEDGEMENTS

This work was supported by the Federations ELESA and IMAG (Grenoble Universities, France – Project "Modèles physiques pour l'étude de la production de la parole et de pathologies des voies aériennes supérieures et de la face »).